\documentclass[epj]{svjour}
\usepackage{graphics}
\usepackage{hyperref}
\usepackage{amsmath,amssymb}
\usepackage{booktabs}
\usepackage{multirow}
\usepackage{booktabs} 
\hypersetup{breaklinks=true}
\usepackage{xcolor}

\begin{document}
\title{Interacting dark sector with quadratic coupling: theoretical and observational viability}


\author{
    Jaelsson S. Lima \thanks{e-mail: \texttt{jaelsson.slima@gmail.com} (corresponding author)} \inst{1} \and
    Rodrigo von Marttens\thanks{e-mail: \texttt{rodrigovonmarttens@gmail.com}} \inst{2,3} \and
    Luciano Casarini\thanks{e-mail:  \texttt{lcasarini@academico.ufs.br}} \inst{1}
}

\institute{
    Departamento de Física, Universidade Federal de Sergipe, São Cristóvão, SE,  49107-230, Brazil \and
    Instituto de Física, Universidade Federal da Bahia, Salvador, BA, 40170-155, Brazil \and
    PPGCosmo, Universidade Federal do Espírito Santo, Vitória, ES, 29075-910, Brazil
}

\date{Received: date / Revised version: date}
%
\abstract{
Models proposing a non-gravitational interaction between dark energy (DE) and dark matter (CDM) have been extensively studied as alternatives to the standard cosmological model. A common approach to describing the DE-CDM coupling assumes it to be linearly proportional to the dark energy density. In this work, we consider the model with interaction term $Q=3H\gamma{\rho_{x}^{2}}/{(\rho_{c}+\rho_{x})}$. We show that for positive values of $\gamma$ this model predicts a future violation of the Weak Energy Condition (WEC) for the dark matter component, and for a specific range of negative values of $\gamma$ the CDM energy density can be negative in the past. We perform a parameter selection analysis for this model using data from Type Ia supernovae from the Pantheon sample, $H(z)$ measurements from the Cosmic Chronometers sample, Baryon Acoustic Oscillations from the DESI survey, and Cosmic Microwave Background data from the Planck combined with the Hubble constant $H_0$ prior. Imposing a prior to ensure that the WEC is not violated, our model is consistent with $\Lambda$CDM in 2$\sigma$ C.L., yet exhibits a preference for smaller values of $\sigma_8$, alleviating the $\sigma_8$ tension between the CMB results from Planck 2018 and the weak gravitational lensing observations from the KiDS-1000 cosmic shear survey. 
%
%
\keywords{{Cosmology} \and {Dark energy} \and {Dark matter} \and {Interacting model} \and {Cosmological parameters}}
} 

\maketitle
%
%

\section{Introduction}\label{sec:I}
In the face of the discovery of the accelerated expansion of the Universe from observations of Type Ia Supernovae (SNe Ia)  \cite{Riess_1998,Perlmutter-1998,Perlmutter_1999}, a new model of the Universe comes into effect, containing characteristics of negative pressure energy content, referred to in the literature as dark energy (DE) \cite{LIMA2004}. Despite the great observational success of the standard $\Lambda$CDM model, where CDM stands for cold dark matter, it presents some shortcomings, such as the cosmological constant problem, the cosmic coincidence, and the Hubble constant tension. Therefore several models or approaches are attempting to obtain a better description of the nature of dark energy, which could be theoretically associated with quantum vacuum states. However, the cosmological constant still stands as the simplest candidate to be the component of dark energy \cite{Ellis_2011,Sami-2009,Marttens-2014,Weinberg1989,Zlatev1999,VONMARTTENS2019,Peebles2003,Hou_2017,Capozziello24,Bamba2012,Pacif2020}.

In this work, we focus on a class of interacting models commonly referred to in the literature as interacting dark energy models (IDEM), with particular attention to the model designated as IDEM 2 in Ref.~\cite{VONMARTTENS2019}. These models depart from the standard assumption of dark matter (DM) and DE as independent components by introducing a non-gravitational interaction, resulting in an energy exchange between them. Such interacting models are primarily phenomenologically motivated, reflecting our limited understanding of the fundamental physics underlying a coupling term \cite{VONMARTTENS2019,vonMarttens2020v2,vonMarttens_2023}.

A common choice for a IDEM assume that the interaction term $Q$ is linearly proportional to the DE density, i.e., $Q = 3\gamma H \rho_x$, where $H$ and $\gamma$ represent the Hubble and interaction parameters, respectively (see, e.g., Refs.~\cite{VONMARTTENS2019,DiValentino:2019ffd,PhysRevD.105.123506,KUMAR2021100862,PhysRevD.101.063502}). However, as demonstrated in Ref.~\cite{vonMarttens2020v2}, this class of models can exhibit non-physical behavior for certain parameter ranges, specifically predicting negative matter densities, thereby violating the Weak Energy Condition (WEC) \cite{NGUYEN2024116669}. In this work, we consider a model in which the interaction term is given by $Q \propto\rho_x^2/\left(\rho_c + \rho_x\right),$ deviating from the conventional linear interaction forms. The quadratic dependence on $\rho_x$ introduces a richer dynamical evolution, reflecting a more complex interplay between DE and CDM. Nevertheless, similar to the linear case, we show that this model also presents non-physical behavior for specific parameter values, with implications for the Universe's background evolution and the growth of cosmic structures \cite{Rowland_2008,MISHRA2023}.

This paper is organized as follows: In Section \ref{sec:II}, we introduce the background dynamics for the cosmological model in analysis. In Section \ref{sec:Methodology} we detail the observational data used, the methodology, and the statistical analysis involved. 
The statistical analysis is based on MontePython~\footnote{The documentation for the code is available at \url{https://github.com/brinckmann/montepython_public/}.} and a suitable modified version of Cosmic Linear Anisotropy Solving System (CLASS)~\footnote{The documentation for the code is available at \url{https://lesgourg.github.io/class_public/class.html}.} codes.
In Section \ref{sec:Results}, we discuss the main results obtained. Finally, we present our conclusions in Section \ref{sec:Conclusions}.

\section{Background Dynamics}\label{sec:II}
Let's consider the energy balance equations, without fixing the state equation parameter at $\omega_x=-1$, but leaving it as a constant parameter. Thus, the energy conservation law for baryons, CDM, and DE are given, respectively, by 
\begin{align}
\dot{\rho}_b+3 H \rho_b&=0\,, \label{benergy} \\ 
\dot{\rho}_{c}+3 H \rho_{c}&=-Q\,, \label{cdmenergy} \\
\dot{\rho}_x+3 H \rho_x\left(1+\omega_x\right)&=Q\,, \label{deenergy}
\end{align}
where $\rho_c$, $\rho_x$, and $\rho_b$ are respectively the densities of CDM, DE, and conserved baryons. The term $\omega_x$ is the DE equation-of-state (EoS) parameter.  In this work, we are interested in the energy exchange between DM and DE described by the IDEM 2 in Ref.~\cite{VONMARTTENS2019}, which is characterized by the interaction term given by 
\begin{align}
Q=3H\gamma\dfrac{\rho_{x}^{2}}{\rho_{c}+\rho_{x}},
\label{q2}
\end{align}
where $H$ and $\gamma$ are respectively the Hubble factor and interaction parameter.

The background solution for the densities of dark matter and dark energy are derived from the resolutions of equations \eqref{cdmenergy} and \eqref{deenergy} and are expressed as follows:
\begin{align}\label{rhoc-c}
\rho_{c}&={\frac{3H_0^2}{8\pi G}}\  a^{-\frac{3\left(\omega_{x}+\omega_{x}^{2}+\gamma\right)}{\omega_{x}+\gamma}}\nonumber \\
 & \times \left[\frac{\Omega_{x 0}\left(\omega_{x}+\gamma\right)+\left(\Omega_{c 0} \omega_{x}-\Omega_{x 0} \gamma\right) a^{3 \omega_{x}}}{\left(\Omega_{c 0}+\Omega_{x 0}\right) \omega_{x}}\right]^{-\frac{\gamma}{\omega_{x+\gamma}}}\nonumber \\
 & \times\left[\Omega_{\mathrm{c0}} a^{3 \omega_{x}}+\Omega_{x 0} \frac{\gamma}{\omega_{x}}\left(1-a^{3 \omega_{x}}\right)\right]
\end{align}
and
\begin{align}\label{rhox2}
 \rho_{x}&={\frac{3H_0^2}{8\pi G}\Omega_{x0}}\ a^{-\frac{3\left(\omega_{x}+\omega_{x}^{2}+\gamma\right)}{\omega_{x}+\gamma}}\nonumber \\
 & \times \left[\frac{\Omega_{x 0}\left(\omega_{x}+\gamma\right)+\left(\Omega_{c 0} \omega_{x}-\Omega_{x 0} \gamma\right) a^{3 \omega_{x}}}{\left(\Omega_{c 0}+\Omega_{x 0}\right) \omega_{x}}\right]^{-\frac{\gamma}{\omega_{x+\gamma}}}\,
\end{align}
where $\Omega_{x0}$ and $\Omega_{c0}$ are the current DE and DM density parameters.
It is possible to express it in a unified way,
\begin{align}\label{ra}
r\left(a\right)=r_{0}\,a^{3\omega_{x}}+\dfrac{\gamma}{\omega_{x}}\left(1-a^{3\omega_{x}}\right)\,,
\end{align}
where $r(a)$ is defined as the ratio between the dark matter density and the dark energy density, $r(a)\equiv \rho_c/\rho_x$. In this case, when $\omega_x <0$, and in the limit where the scale factor $a(t)$  tends to infinity, $r(a)$  approaches a constant value, which differs from the standard $\Lambda$CDM model that tends to zero. First-order Newtonian gauge was used in the perturbative contribution, as presented in Ref. \cite{VONMARTTENS2019}.

We are interested in determining the regions where the densities $\rho_x$ and $\rho_c$ are greater than or equal to zero. Specifically, we want to find the allowed regions for the DE and DM densities during their cosmic evolution, as only these have physical significance. Given the definition $r(a)=\rho_c/\rho_x$, it is sufficient to evaluate only one of the pairs: $\rho_x, \rho_c$ or $\rho_x, r(a)$ or $\rho_c, r(a)$. To ensure that the curves have values greater than or equal to zero, from the initial scale factor $a=0$ to $a\rightarrow \infty$, it is suitable to introduce a variable transformation of the type:
\begin{equation}\label{eq:transformacao}
 y=\frac{2}{\pi}\arctan(\bar{y}), 
\end{equation}
where $\bar{y}$ can be: $\rho_{c}$, $\rho_{x}$, $r$, or $a$. The x-axis, used for the scale factor, assumes the finite interval $x=[0,1]$, while y, being used for $\rho_{x}$, or $r$, assumes the finite interval $y=[-1,1]$.

\begin{figure}[!htbp]\centering
\resizebox{0.5\textwidth}{!}{\includegraphics{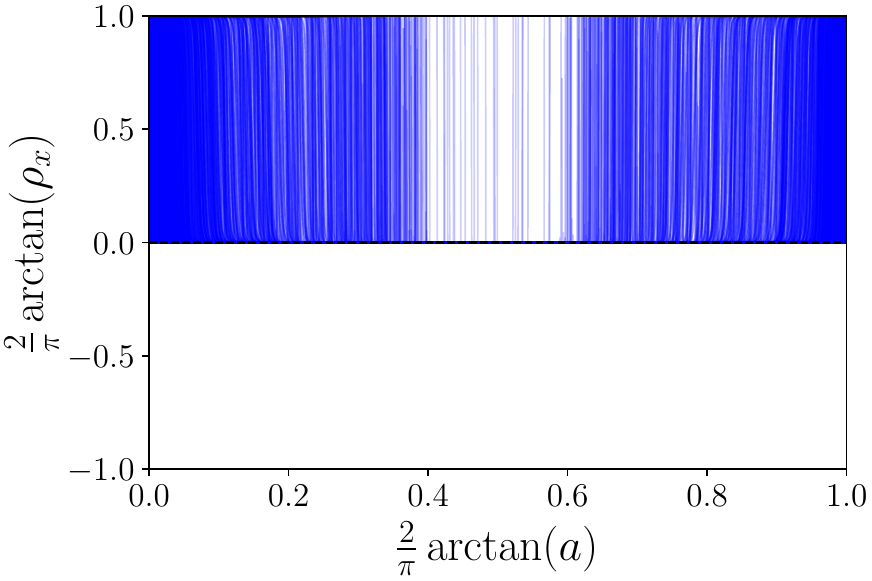}}
\caption{Computation with 30000 of curves of $\rho_{x}(a)$. The blue lines represent curves with values $\rho_x(a)\geq0$. There were no curves for $\rho_x(a)<0$. The graph represents the largest possible scale for $\rho_x$ and $a$ ($\rho_x=[-\infty,\infty]$, $a=[0,\infty]$), with the parameters $\omega_x, ~ \gamma,~\Omega_{m0}$, and $h$ randomly varying between the intervals of $\omega_x = [-2,2]$, $\gamma = [-2,2]$, $\Omega_{m0}=[-1,1]$, and $h=[0.5,1]$, respectively.\label{fig:rhox_vs_a}}
\end{figure}

\begin{figure} [!htbp]\centering
\resizebox{0.5\textwidth}{!}{\includegraphics{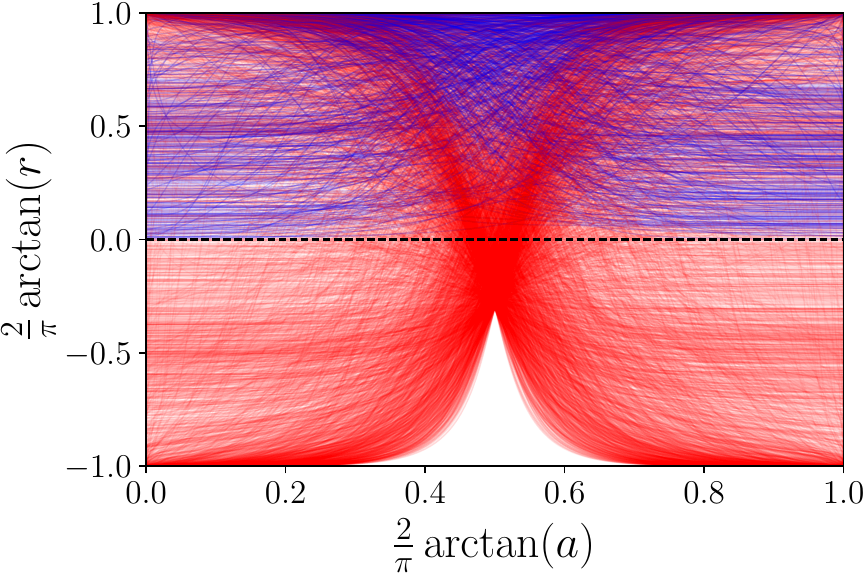}}
\caption{Computation with 5000 of curves of $r(a)$. The blue lines represent curves with values $r(a)\geq0$, and the red lines represent curves with values that, at some point, have $r<0$. The graph represents the largest possible scale for $r$ and $a$ ($r=[-\infty,\infty]$, $a=[0,\infty]$), with the parameters $\omega_x, ~ \gamma$, and $~\Omega_{m0}$ randomly varying between the intervals of $\omega_x = [-2,2]$, $\gamma = [-2,2]$, and $\Omega_{m0}=[-1,1]$, respectively.\label{fig:r(a)_vs_a} }
\end{figure}

\begin{figure}[!htbp]\centering
 \resizebox{0.5\textwidth}{!}{\includegraphics{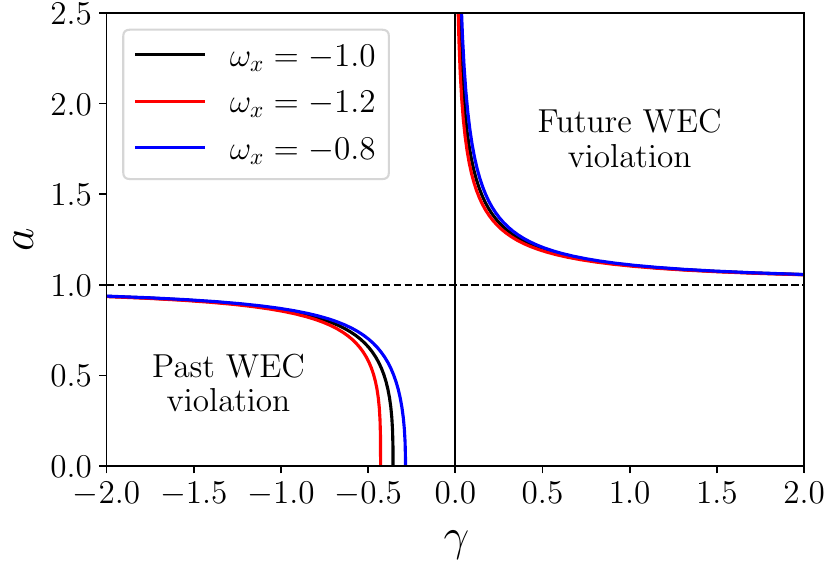}}%
\caption{Plot of the scale factor $a$ that violates the WEC as a function of the interaction parameter $ \gamma$. On the top right is shown the WEC violation in the future, while on the bottom left is shown the WEC violation in the past. The black line corresponds to $\omega_{x}=-1$, while the blue line corresponds to $\omega_{x}=-0.8$, and the red line corresponds to $\omega_{x}=-1.2$.
Here, the dark energy density parameter was fixed at $\Omega_{x0}=0.7$. \label{fig:IDEM2_a_vs_gamma} }
\end{figure}

To ensure that equation \eqref{rhox2} satisfies $\rho_x\geq0$, in Figure \ref{fig:rhox_vs_a}, was plotted $({2}/{\pi})\arctan(\rho_x)$ versus $({2}/{\pi})\arctan(a)$ from a computation with 30000 of generated curves with the parameters $\omega_x, ~ \gamma,~\Omega_{m0}$, and $h$ (where $h$ is $H/100$ km$/$s$/$Mpc) randomly varying between the intervals of $\omega_x = [-2,2]$, $\gamma = [-2,2]$, $\Omega_{m0}=[-1,1]$, and $h=[0.5,1]$, respectively. 
Meanwhile, Figure \ref{fig:r(a)_vs_a} shows the plot of $({2}/{\pi})\arctan(r)$ versus $({2}/{\pi})\arctan(a)$, built with the generation of 5000 curves, with the same random choice of the parameters $\omega_x, ~ \gamma,~\Omega_{m0}$. 
In this figure, the dark matter density can assume negative values ($\rho_{c}<0$) for certain time intervals (in the past or in the future), thereby violating the WEC.
From equation \eqref{rhoc-c}, it is possible to assess the conditions leading to the WEC violation. Thus, we can divide the analysis into two distinct periods: the past and the future, so that it can be observed that the energy density of matter assumes negative values at
\begin{align} \label{eq:a}
a=\left[\frac{-\gamma\,\Omega_{x0}}{ \omega_x\Omega_{c0} - \gamma \Omega_{x0}}\right]^{\frac{1}{3\,\omega_x}  }.
\end{align}
Assuming that the value of the equation of state parameter is $\omega_x<0$, we straightly obtain the solution conditions for the equation \eqref{eq:a} in the intervals $0<a<1$ (in the past) and $a>1$ (in the future).
To prevent the dark matter energy density from becoming negative in the early moments of the universe, the interaction parameter must satisfy the following condition 
\begin{align}\label{eq:gamma}
\gamma >-\left|\omega_x\right| \frac{\Omega_{c0}}{\Omega_{x0}} .
\end{align}
However, it is likely that the density of dark matter, $\rho_c$, will be negative in the future, unless,
\begin{align}\label{eq:gamma2}
\gamma \leq 0,
\end{align}
that is independent of the value $\omega_x$. 
The equations \eqref{eq:gamma} and \eqref{eq:gamma2}  define the range in which the model is physically well-defined, regardless of the timescale. 
In Figure \ref{fig:IDEM2_a_vs_gamma}, the scale factor is depicted for the WEC violation as a function of the interaction parameter $\gamma$.

\begin{figure*}
    \resizebox{0.32\textwidth}{!}{\includegraphics{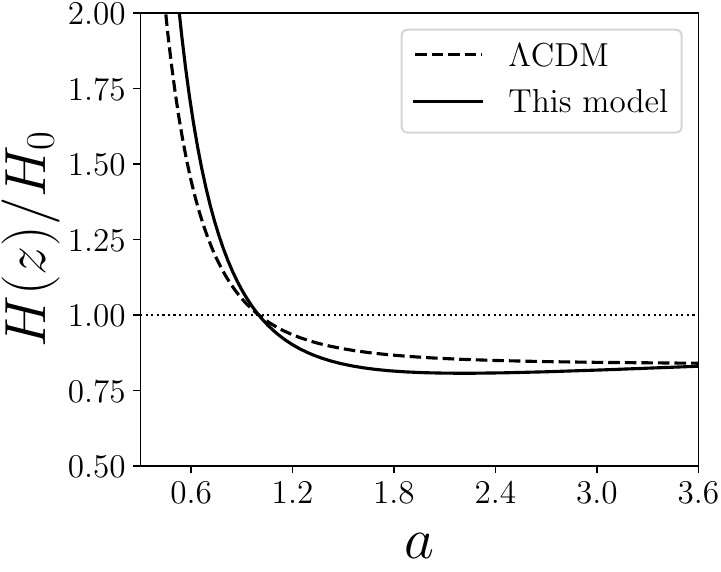}}
    \resizebox{0.31\textwidth}{!}{\includegraphics{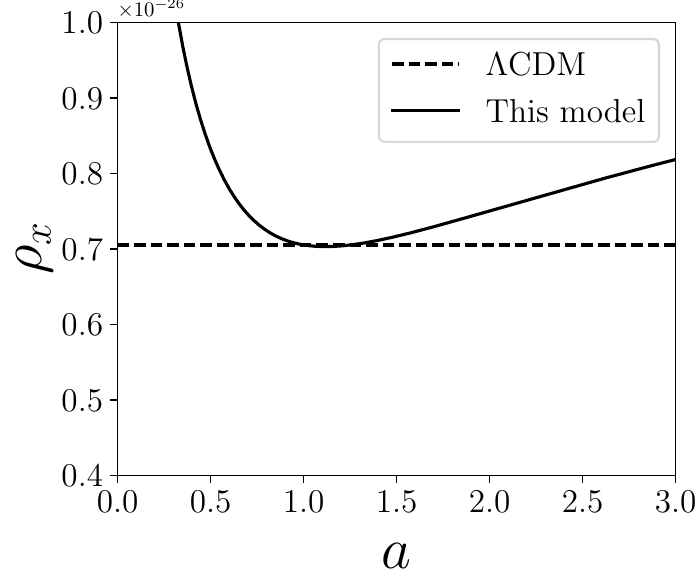}}
    \resizebox{0.33\textwidth}{!}{\includegraphics{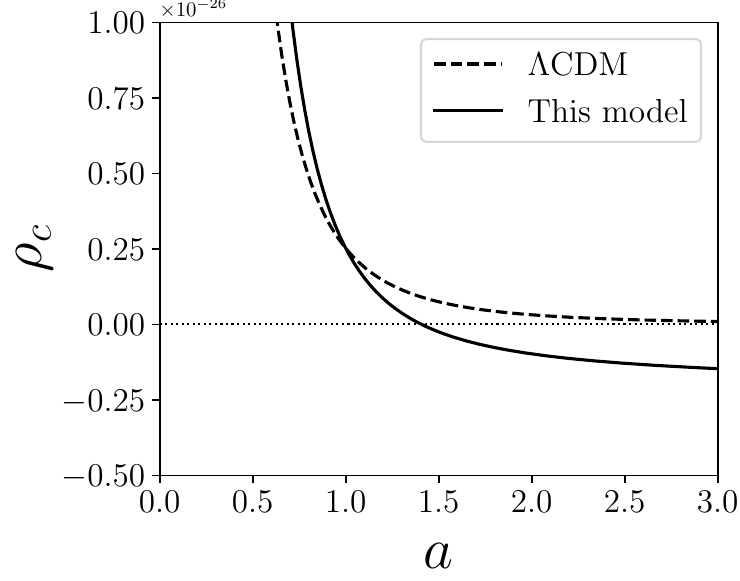}}
\caption{Background solutions of this model and the
$\Lambda$CDM model as a function of the scale factor $a$. Left panel: Shows the Hubble function $H(z)/H_0$. Central panel: Dark energy density. Right panel: Dark matter density. In all panels, the solid line corresponds to the model, with  $\omega_x=-1$, $\Omega_{x0}=0.7$, $\gamma=0.2$ and $h=0.7324$, and the dashed lines correspond to the $\Lambda$CDM model, with the same values of $\Omega_{x0}$ and $h$.
\label{fig:Hz}}
\end{figure*}

Let's consider, for example, $\omega_x=-1$, $\Omega_{x0}=0.7$ and $\gamma=0.2$ $\rho_c$ becomes negative with scale factor $a\approx 1.4$, where the critical density is $\rho_{c r 0}=\frac{3 H_0^2}{8 \pi G}=1.87798 \times 10^{-26} h^2 \frac{\mathrm{kg}}{\mathrm{m}^3}$, with \( h = 0.7324 \) \cite{Riess:2016jrr}. Figure \ref{fig:Hz} shows how the universe evolves with the Hubble function, DE, and matter densities as a function of the scale factor.

\section{Methodology}\label{sec:Methodology}
This section briefly describes the observational datasets and the statistical analysis methodology.

\subsection{Observational Data} \label{sec:data}
Here we will present the observational datasets.
\begin{itemize}

\item \textbf{Type Ia Supernovae (SNe Ia):}
SNe Ia are considered standard candles in astronomy and have played a pivotal role in observing the Universe's accelerated expansion \cite{Planck-2018vi,Riess_1998,Perlmutter-1998,Perlmutter_1999}. In general, the apparent magnitude and luminosity distance are related by $m=5\log \left(D_L\right)+25+M\,$, where $M$ is the absolute magnitude. $D_L(z)$ represents the luminosity distance expressed in units of Mpc. In this analysis, $1048$ measurements of apparent magnitudes from Type Ia supernovae were used, referred to as the Pantheon sample\footnote{The Pantheon data is available and can be downloaded from \url{http://www.github.com/dscolnic/Pantheon}.}. This dataset covers the redshift range $0.01<z<2.3$ \cite{Scolnic_2018,Betoule-2014}.

\item \textbf{Current Value of the Hubble Expansion Rate ($H_0$):}
The Hubble expansion rate obtained by \cite{Riess:2016jrr} provides the best estimate of $H_0=(73.24\pm 1.74)\,$km s$^{-1}$ Mpc$^{-1}$, from a set of observations containing more than 600 Cepheids, using both infrared and visible frequencies. This value is independent of the cosmological model.

\item \textbf{Cosmic Chronometers (CC):}
The Cosmic Chronometers are independent data from cosmological models, derived from measurements taken from ancient galaxies. There are $30$ measurements of $H(z)$ in the range $0.07<z<1.965$~\footnote{The CC data has been included in MontePython and is available at \url{https://github.com/brinckmann/montepython_public/tree/3.3/data/cosmic_clocks}.} \cite{Moresco_2016}.

\item \textbf{Baryon Acoustic Oscillations (BAO):}
Baryon Acoustic Oscillations carry information from the pre-decoupling Universe.
The baryon-photon fluid propagates with an acoustic velocity described as follows \cite{Bassett2010-BAO,Eisenstein-2007}:
\begin{align}\label{eq:som}
c_s=\frac{c}{\sqrt{3(1+\mathrm{R})}},
\end{align}
with $\mathrm{R}\equiv\frac{3\rho_b}{4\rho_\gamma}\propto\frac{\Omega_b}{1+z}$.

After decoupling, photons begin to travel with a characteristic scale given by
\begin{eqnarray}
r_s(a_{\rm{drag}})\equiv\int_{0}^{a_{\rm drag}}\dfrac{c_{s}\left(a\right)}{Ha^{2}}\ da\,, \label{rs}
\end{eqnarray}
where $c_s$ represents the speed of sound in the primordial plasma and $a_{\rm drag}$ is the scale factor at the drag time, referring to the moment in the early universe when photons and baryons (protons and neutrons) decoupled \cite{PhysRevD.104.043521}. 
The isotropic BAO measurements are provided through the dimensionless ratio $D_V/ r_s(a_{\rm{drag}})$, where $D_V$ is the geometric mean that combines the scales of the line-of-sight and transverse distances. $D_V$ is expressed by
\cite{Eisenstein_2005,Haridasu_2018} 
\begin{eqnarray}
D_{V}(z)=\left[\left(1+z\right)^{2}D_{A}^{2}(z)\dfrac{cz}{H(z)}\right]^{1/3}\,,
\end{eqnarray}
where $D_A(z)$ is the angular diameter distance. 
Other key distance quantities measured are the transverse comoving distance, $D_M(z)/r_s(a_{\rm{drag}})$, and the Hubble horizon, $D_H(z)/r_s(a_{\rm{drag}})$, where $D_M(z)$ is the comoving angular diameter distance that is associated with $D_L(z)$ ($ D_M(z) = D_L(z)/(1+z) $), and $D_H$ is the Hubble distance ($D_H(z) = c/H(z)$). Here, we consider BAO data from the Dark Energy Spectroscopic Instrument (DESI) survey. The measurements are summarized in Table I of Ref. \cite{desicollaboration2024desi2024vicosmological}
.

\item \textbf{CMB data (Planck):}
The data used here for CMB are measurements from Planck 2018, which include information on temperature, polarization, temperature polarization cross-correlation spectra, and lensing maps reconstruction, Planck (TT, TE, EE+lowE+\textit{lensing}) \cite{planck18-v}.\footnote{For CMB analysis, all Planck likelihood codes and
data can be obtained at \url{https://pla.esac.esa.int/pla/}.}
In these analyses, the standard likelihood codes were considered: (i) COMANDER for the low-$l$ TT spectrum, with spectrum data ranging from $2 \leq l < 30$; (ii) SimAll for the low-$l$ EE spectrum, with spectrum data ranging from $2 \leq l < 30$; (iii) Plik TT, TE, EE for the TT, TE, and EE spectra, covering $30 \leq l \lesssim 2500$ for TT and $30 \leq l \lesssim 2000$ for TE and EE; (iv) lensing power spectrum reconstruction with $8 \leq l \leq 400$. For more details on the likelihoods, see Refs. \cite{planck18-v,Planck-2018vi}.

\end{itemize}

\subsection{Statistical analysis }\label{sec:statistical} 

Statistical analysis is performed using the MontePython code \cite{Brinckmann-2018,BenjaminAudren_2013}, which utilizes the CLASS code \cite{Blas_2011,BenjaminAudren_2013}. In the MontePython code, the Markov Chain Monte Carlo (MCMC) method \cite{Metropolis-1953,Hastings-1970} is used to perform statistical analysis on the input data, comparing it with the theoretical predictions, that are provided by a suitably modified version of the CLASS code, to take into account the cosmology framework described in section \ref{sec:II}. 
To use the $\Lambda$CDM and $\omega$CDM models, no modifications are necessary in the code. However, for other models, such as the interacting models, it is necessary to implement the background equations as well as the perturbed fluid equations with the linear perturbative contribution of the model in the code. 
For all chains, during the MCMC analysis in MontePython, it is required that the Gelman-Rubin convergence parameter satisfies the condition $\hat{R}-1 < 0.01$ \cite{2018arXiv181209384V,gelman1992}. We utilized \texttt{GetDist}\footnote{The documentation for the code is available at \url{https://getdist.readthedocs.io/}.} to analyze and plot the chains \cite{Lewis_2019,2019ascl.soft10018L}. 
It allows for creating contour plots, histograms, parameter correlations, among others, from datasets generated by codes such as MontePython. In this part, using the \texttt{GetDist} library, the samples where the parameter $\gamma$ does not satisfy the conditions of equations (\ref{eq:gamma}) and (\ref{eq:gamma2}) were filtered. These equations establish specific limits for the value of $\gamma$. Thus, only the samples that adhere to these restrictions are considered in the posterior analysis, specifically with the prior applied.

\subsection{Combination of dataset}
We investigate the impact of the physical or WEC prior, in the case of $\gamma \geq 0$, on the estimation of the cosmological parameters of the model under analysis. 
The unmodified CLASS code already automatically resolves past WEC violation.
A Bayesian statistical analysis was conducted with and without the inclusion of such a prior, for five different datasets:
\begin{enumerate}
   \item Background: \textit{Composed of SNe Ia, CC, and  DESI BAO data};
    \item Background+$H_0$: \textit{Composed of SNe Ia, CC, DESI BAO, and $H_0$ data};
    \item Planck: \textit{ Composed of the full CMB data from Planck, combined with Planck TTTEEE+lensing reconstruction};
    \item Background+Planck: \textit{Composed of the combination of Planck+Background};
    \item Background+Planck+$H_0$: \textit{ Composed of the combination of Planck+Background, with $H_0$}.
\end{enumerate}

\section{Results and discussion }\label{sec:Results} 

When we do not include the Planck data in our analysis, we considered the following cosmological parameters: $\mathcal{P}$=$\{\omega_{c}$, $H_0$, $\gamma \}$ where $\omega_{c}$ is the physical cold dark matter density parameter, and the derived parameter: $\mathcal{P^{'}}$=$\{\Omega_{m0}\}$ where $\Omega_{m0}$ represent the current matter density parameter.
Otherwise when we include the data from Planck, the parameters are: $\mathcal{P}$=$\{100\theta{}_{s}$, $n_{s}$, $\ln$$(10^{10}A_{s })$, $\omega_{b}$, $\omega_{c}$, $\tau_{reio}$, $\gamma \}$ where $100\theta_{s}$, $n_{s}$, $\ln(10^{10}A_{s})$, $\omega_{b}$, and $\tau_{reio}$ are respectively the angular size of the sound horizon (scaled by 100), the scalar spectral index, the logarithm of the amplitude of the primordial scalar power spectrum, the physical baryon density, and the optical depth to reionization, and the derived parameters: $\mathcal{P^{'}}$=$\{H_0$, $\Omega_{m0}$, $\sigma_{8}$, $S_8 \}$, where $\sigma_{8}$ and $S_8$ are respectively the standard deviation of the density fluctuation in an 8 $h^{-1}$ Mpc radius sphere, and the structure growth parameters, with the equation of state parameter fixed at $\omega_x=-1$.
The results of our analysis are presented in Tables \ref{tab:idem2_background} and \ref{tab:idem2_planck}, both with and without the WEC prior. 
For the background tests, the physical baryon density parameter was fixed at $\omega_b=\Omega_{b0} h^2=0.022$ \cite{desicollaboration2024desi2024vicosmological}, while in the analyses that include CMB data, $\omega_{b}$ is a free parameter.
\begin{table*}[!htbp]
\centering
\caption{Statistical analysis results at $1\sigma$ C.L., considering the analyses without and with the WEC prior, using the datasets: Background and Background+$H_0$.\label{tab:idem2_background}}
\begin{tabular}{lcccc}
\toprule \toprule
\multirow{2}{*}{Parameter} & \multicolumn{2}{c}{Background} & \multicolumn{2}{c}{Background+$H_0$} \\
\cmidrule(lr){2-3} \cmidrule(lr){4-5}
                         & No prior & WEC prior  & No prior  & WEC prior \\
\midrule
$\gamma                    $ & $0.01^{+0.12}_{-0.12}$  & $> -0.126 $ & $0.10^{+0.11}_{-0.11}$ & $>-0.092$ \\
$\omega_{c}                $ & $0.116^{+0.019}_{-0.015}$ & $0.1293^{+0.0091}_{-0.0110}$ & $0.104^{+0.020}_{-0.018}$ & $0.1292^{+0.0082}_{-0.0092}$ \\
$H_0                       $ & $69.0^{+1.2}_{-1.2}$     & $68.13^{+0.80}_{-0.80}$      & $70.2^{+1.0}_{-1.0}$     & $68.99^{+0.65}_{-0.59}$ \\ \midrule
$\Omega_{m0}               $ & $0.291^{+0.047}_{-0.040}$ & $0.326^{+0.021}_{-0.030}$ & $0.256^{+0.044}_{-0.044}$ & $0.318^{+0.018}_{-0.022}$ \\
\bottomrule \bottomrule
\end{tabular}
\end{table*}

\begin{table*}[!htbp]
    \centering
      \caption{Statistical analysis results at $1\sigma$ C.L., considering the analyses without and with the WEC prior, using the datasets: Planck, Planck+Background, and Planck+Background+$H_0$. \label{tab:idem2_planck}}
 \begin{tabular} {l c c c c c c}
  \\ \toprule \toprule
\multirow{2}{*}{Parameter} & \multicolumn{2}{c}{Planck} & \multicolumn{2}{c}{Planck+Background} & \multicolumn{2}{c}{Planck+Background+$H_0$}\\
\cmidrule(lr){2-3} \cmidrule(lr){4-5} \cmidrule(lr){6-7}
                         & No prior  & WEC prior  & No prior  & WEC prior  & No prior  & WEC prior \\
\midrule
$\gamma                    $ & $0.08^{+0.14}_{-0.11}      $ & $>-0.117 $ & $0.077^{+0.078}_{-0.078}   $ & $>-0.058  $ & $0.107^{+0.067}_{-0.079}            $ & $>-0.0161  $\\

{$100\theta{}_{s }$}    & $1.04191^{+0.00030}_{-0.00030}         $ & $1.04187^{+0.00030}_{-0.00030}        $ & $1.04206^{+0.00030 }_{-0.00030 }       $ & $1.042213^{+0.00034}_{-0.00034}$ & $1.04212^{+0.00027}_{-0.00027}        $ & $1.04215^{+0.00024}_{-0.00024}        $\\

{$\ln10^{10}A_{s }$}     & $3.045^{+0.015}_{-0.015}   $ & $3.044^{+0.014}_{-0.014}            $ & $3.051^{+0.014 }_{-0.014}           $ & $3.052^{+0.015}_{-0.015}   $ & $3.053^{+0.014}_{-0.014}             $ & $3.056^{+0.012}_{-0.012}   $\\

{$n_{s }$}     & $0.9654^{+0.0041}_{-0.0041}          $ & $0.9647^{+0.0040}_{-0.0040}          $ & $0.9696^{+0.0035}_{-0.0035}           $ & $0.9707^{+0.0035}_{-0.0031}          $ & $0.9704^{+0.0036}_{-0.0036}$ & $0.9722^{+0.0037}_{-0.0037}           $\\

$\omega_{b}                $ & $0.02238^{+0.00014}_{-0.00014}        $ & $0.02236^{+0.00014}_{-0.00014}        $ & $0.02250^{+0.00013}_{-0.00013}        $ & $0.02250^{+0.00013}_{-0.00013}         $ & $0.02254^{+0.00013}_{-0.00013}        $ & $0.02263^{+0.00013}_{-0.00013}        $\\

$\omega_{c}                $ & $0.104^{+0.022}_{-0.022}   $ & $0.1287^{+0.0054}_{-0.0054 }          $ & $0.103^{+0.018}_{-0.012 }          $ & $0.1236^{+0.0040}_{-0.0046}$ & $0.097^{+0.017}_{-0.012}   $ & $0.1191^{+0.0012}_{-0.0014}$\\

$\tau_{reio}$ & $0.0544^{+0.0076}_{-0.0076}         $ & $0.0539^{+0.0070}_{-0.0070}         $ & $0.0588^{+0.0072}_{-0.0072}$ & $0.0592^{+0.0076}_{-0.0076}          $ & $0.0504^{+0.0074}_{-0.0074}         $ & $0.0628^{+0.0070}_{-0.0064}           $\\\midrule


$H_0                       $ & $68.2^{+1.5}_{-1.3}        $ & $66.70^{+0.70}_{-0.70}              $ & $68.97^{+0.79 }_{-0.79 }             $ & $68.00^{+0.41}_{-0.41}              $ & $69.38^{+0.65}_{-0.75}     $ & $68.53^{+0.36}_{-0.36}     $\\

$\Omega_{m0}               $ & $0.275^{+0.058}_{-0.066}   $ & $0.347^{+0.020}_{-0.027}             $ & $0.267^{+0.042}_{-0.032}             $ & $0.318^{+0.011}_{-0.012}   $ & $0.251^{+0.039}_{-0.031}   $ & $0.3033^{+0.0052}_{-0.0052}$\\

$\sigma_{8}   $ & $0.93^{+0.14}_{-0.17}      $ & $0.763^{+0.036}_{-0.029}   $ & $0.905^{+0.052}_{-0.130}     $ & $0.782^{+0.017}_{-0.017}   $ & $0.947^{+0.066}_{-0.140}    $ & $0.8001^{+0.0081}_{-0.0065}   $\\

$S_8                       $ & $0.869^{+0.041}_{-0.057}   $ & $0.819^{+0.015}_{-0.015}         $ & $0.845^{+0.022}_{-0.044}   $ & $0.8038^{+0.0098}_{-0.0088}   $ & $0.857^{+0.026}_{-0.048}   $ & $0.804^{+0.010}_{-0.009}   $\\
\bottomrule \bottomrule
\end{tabular}
\end{table*}

\begin{figure*}[!htbp]\centering
\resizebox{1\textwidth}{!}{\includegraphics{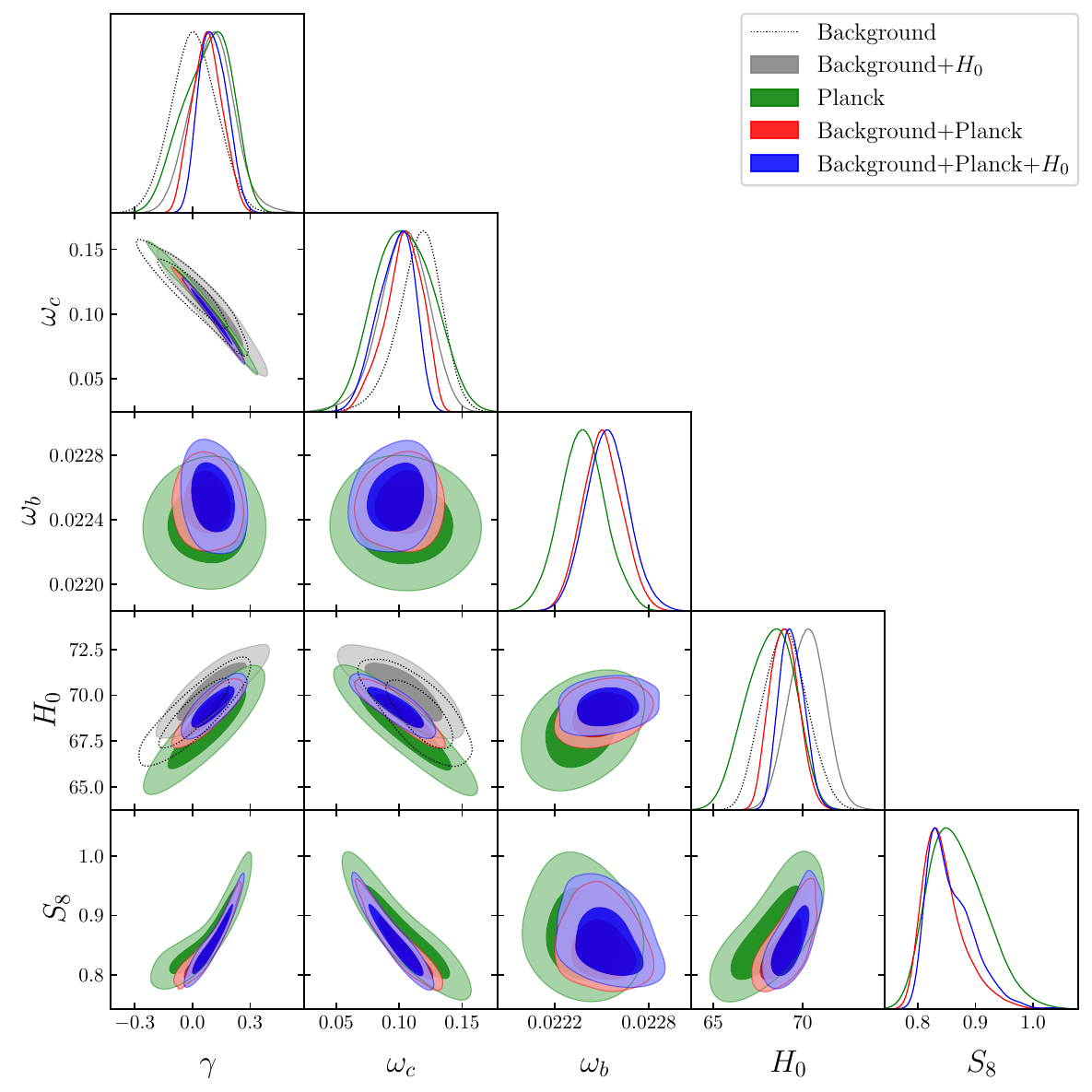}}
\caption{Triangular plot with the free cosmological parameters, considering the analyses without WEC prior, using the datasets: Background, Background+$H_0$, Planck, Background+Planck and Background+Planck+$H_0$. \label{fig:idem2_retangular}}
\end{figure*}
\begin{figure*}[!htbp]\centering
\resizebox{1\textwidth}{!}{\includegraphics{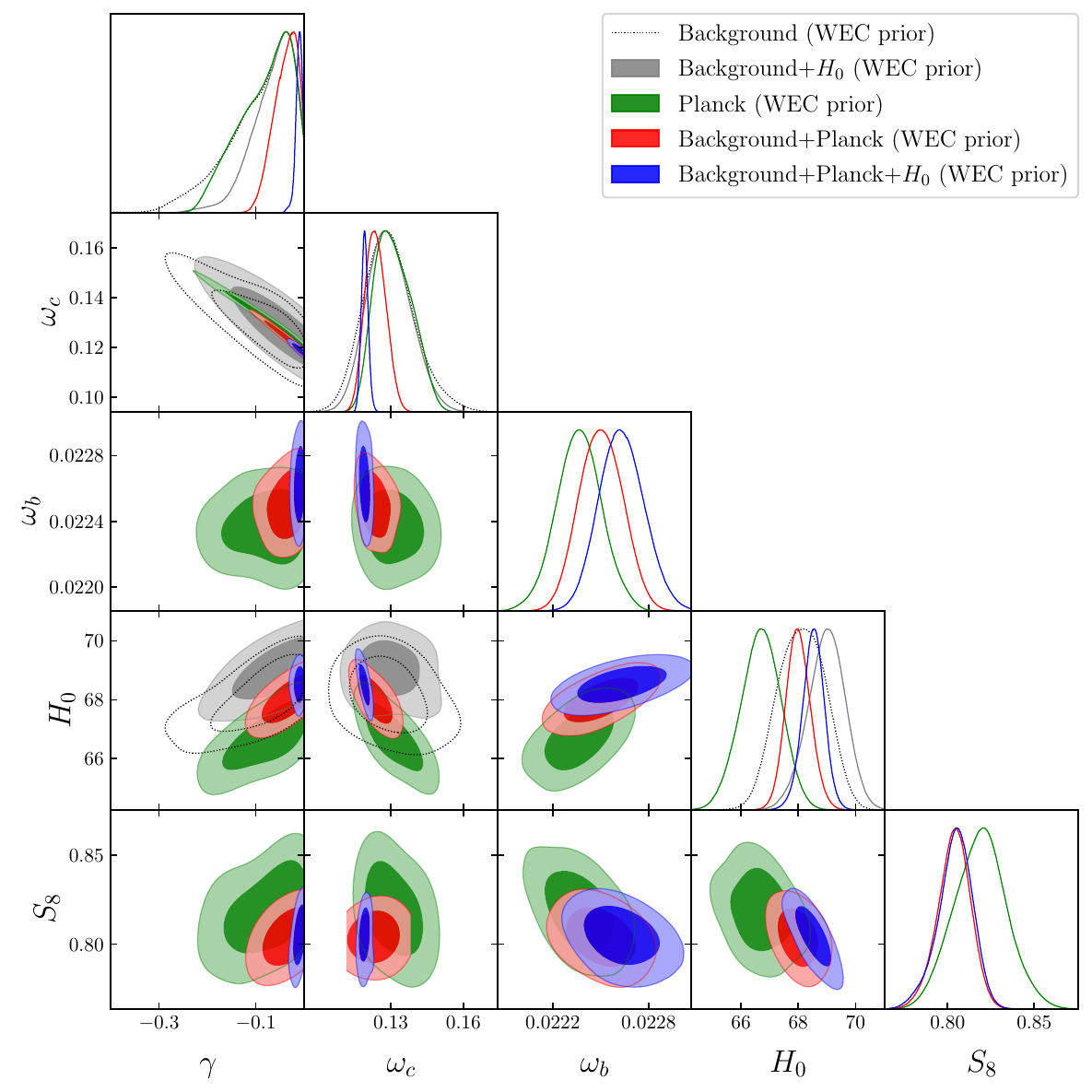}}
\caption{Triangular plot with the free cosmological parameters, considering the analyses with WEC prior, using the datasets: Background, Background+$H_0$, Planck, Background+Planck and Background+Planck+$H_0$. \label{fig:idem2_retangular_wec}}
\end{figure*}

In Figure \ref{fig:idem2_retangular}, the contour curves and posterior distributions without the implementation of the WEC constraint in the background solutions are shown. In Figure \ref{fig:idem2_retangular_wec}, the contour curves and posterior distributions taking into account the WEC constraint are presented. 
Both figures display contour regions at  $1\sigma$ and $2\sigma$ confidence level (C.L.), respectively. 
The confidence regions for $H_0$, $\Omega_{m0}$ and $\sigma_8$ are less restrictive without the WEC prior and more restrictive with it, being consistent with the Planck results \cite{Planck-2018vi} at 2.3$\sigma$ C.L..
Figures \ref{fig:idem2_retangular} and \ref{fig:idem2_retangular_wec} show an anticorrelation between the interaction parameter $\gamma$ and the parameter $\omega_{c}$, and exhibits a correlation between the interaction parameter $\gamma$ and the parameters $\sigma_8$, $S_8$ and $H_0$. 
The same trends accompany other models that exhibit similar coupling physics and are explained physically in \cite{VOMMARTTENS2017114,VONMARTTENS2019}.

%
The triangular plot in Figures \ref{fig:idem2_retangular} and \ref{fig:idem2_retangular_wec} show
the posterior results for the interaction parameter $\gamma$ are presented in two scenarios: without the use of a prior and with the application of the WEC prior. 
For the case without prior, in all analyses, the mean value of $\gamma$ is slightly greater than zero, which means that the WEC will inevitably be violated in the future.
In contrast, when the WEC prior is taken into account, the mean value of $\gamma$ is always negative, satisfying the WEC. 

The results with WEC prior at $1\sigma$ C.L. for $\gamma$ do not satisfy the standard $\Lambda$CDM model. In all cases, the $\Lambda$CDM limit ($\gamma=0$) is satisfied within the $2\sigma$ C.L..

The triangular plot in Figures \ref{fig:idem2_retangular} and \ref{fig:idem2_retangular_wec} show the posteriors for $H_0$.
Both for the case with the WEC prior and for the one without it, when the datasets Background, Background+$H_0$, and Planck are used there are weaker constraints on the Hubble constant, with a best-fit mean around $H_0 \approx 67.9-69.1$ km s$^{-1}$ Mpc$^{-1}$. By combining Background data with Planck data, the constraints are improved. 
In \autoref{sec:appA}, we show our analysis for the $\Lambda$CDM model ($\gamma=0$) with the same datasets for the purpose of comparison. The $H_0$ tension between the analysis that includes the Planck data and the analysis that includes the background data only is reproduced. The interacting model is consistent with $\Lambda$CDM in 2$\sigma$ C.L., and we can not appreciate a significant decreasing of the $H_0$ tension (see Tabs. \ref{tab:lcdm_background} and \ref{tab:lcdm_planck}).    
\begin{figure}[!htbp]\centering
       \resizebox{0.5\textwidth}{!}{\includegraphics{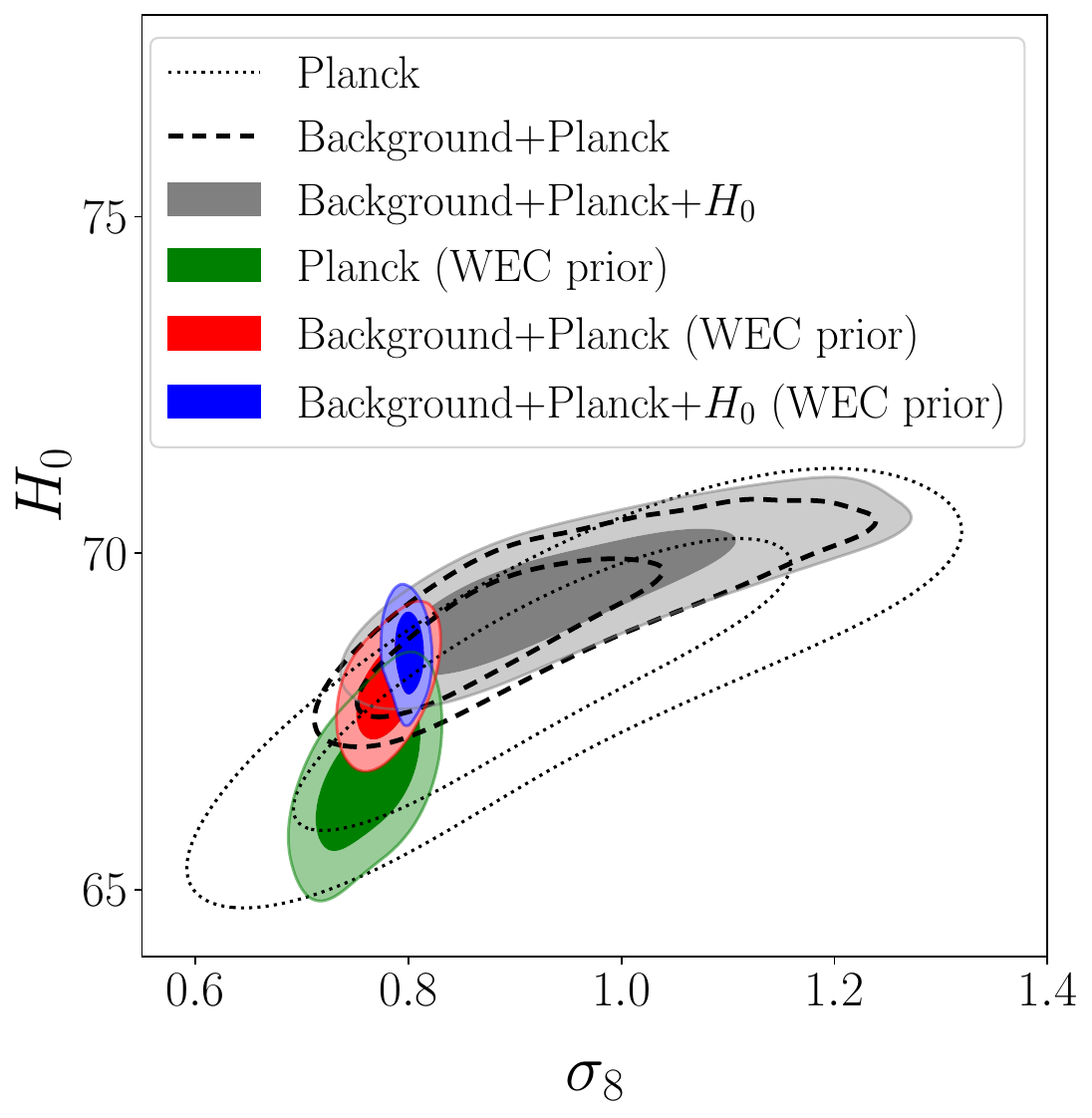}}
\caption{ Plot of the parameter $\sigma_8$ versus the Hubble parameter $H_0$ without the WEC prior and with the WEC prior.\label{fig:plots_sigma8_pert}}
\end{figure}

In Figure \ref{fig:plots_sigma8_pert}, the plot for the plane $\sigma_{8}$ - $H_0$ is shown. 
In the case where the WEC prior was not used the constraints are weaker, while in the case the prior was adopted, the constraints are more stringent, and the estimates tend toward to lower value of $\sigma_8$. The results of the analyses with the WEC prior yield average values of approximately 67.7 km s$^{-1}$ Mpc$^{-1}$ and 0.78 for $H_0$ and $\sigma_8$, respectively. The preference for lower values of $\sigma_8$ alleviates the $\sigma_8$ tension between the primary CMB results from Planck 2018 \cite{Planck-2018vi} and the weak gravitational lensing observations from the KiDS-1000 cosmic shear survey \cite{Catherine_2021}.

\begin{figure*}[!htbp]\centering
        \resizebox{0.46\textwidth}{!}{\includegraphics{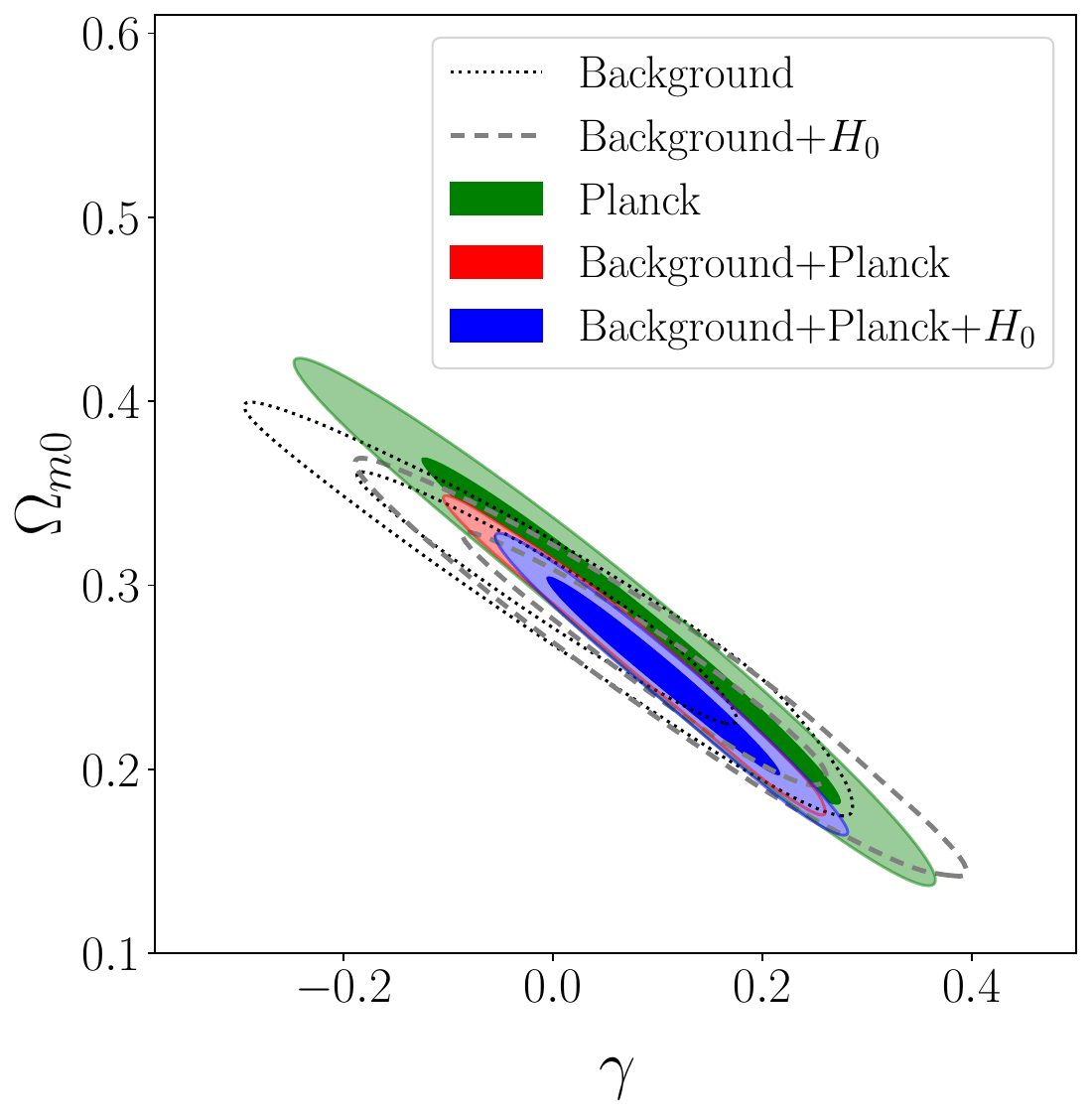}}
        \resizebox{0.47\textwidth}{!}{\includegraphics{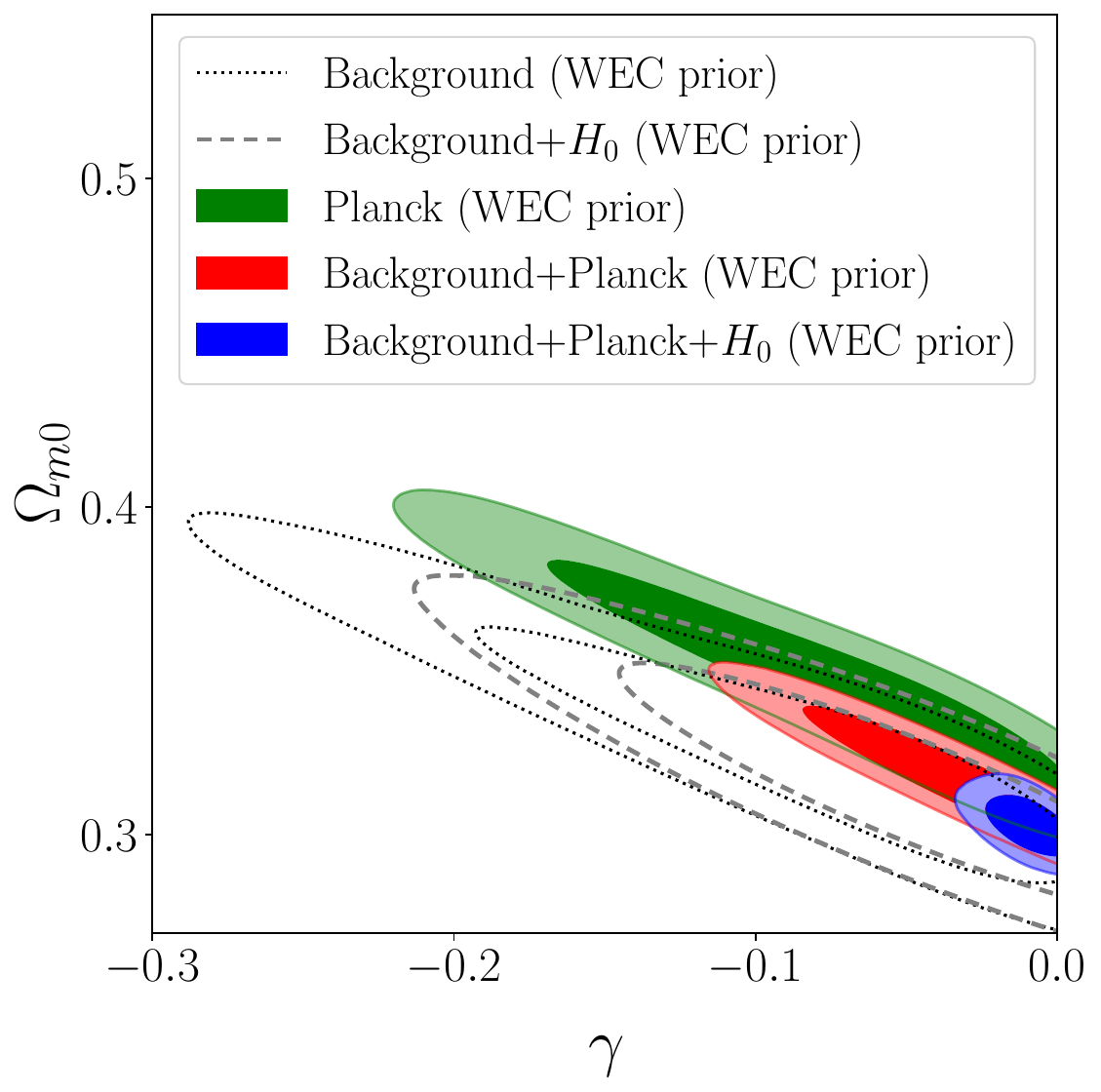}}
\caption{Plot of the interaction parameter $\gamma$ versus the matter density parameter $\Omega_{m0}$ without the WEC prior (left panel) and with the WEC prior (right panel). \label{fig:plots_2d}}
\end{figure*}

Figure \ref{fig:plots_2d} shows the plot of the interaction parameter $\gamma$ versus the matter density parameter $\Omega_{m0}$ with and without the WEC prior. In the case without the WEC prior, all datasets have weak constraints on the parameters $\gamma$ and $\Omega_{m0}$, which appear to improve when the Background and $H_0$ data are combined with the Planck data. 
The inclusion of the WEC prior significantly improves the constraints of the parameters, resulting in higher values of $\Omega_{m0}$, which helps to alleviate the $S_8$ tension between the Planck CMB results and the estimates from the KiDS cosmic shear survey,  as discussed in Refs. \cite{Planck-2018vi,Catherine_2021,Joudaki_2017,10.1093/mnras/stx2820}.


\begin{figure}[!htbp]\centering
        \resizebox{0.5\textwidth}{!}{\includegraphics{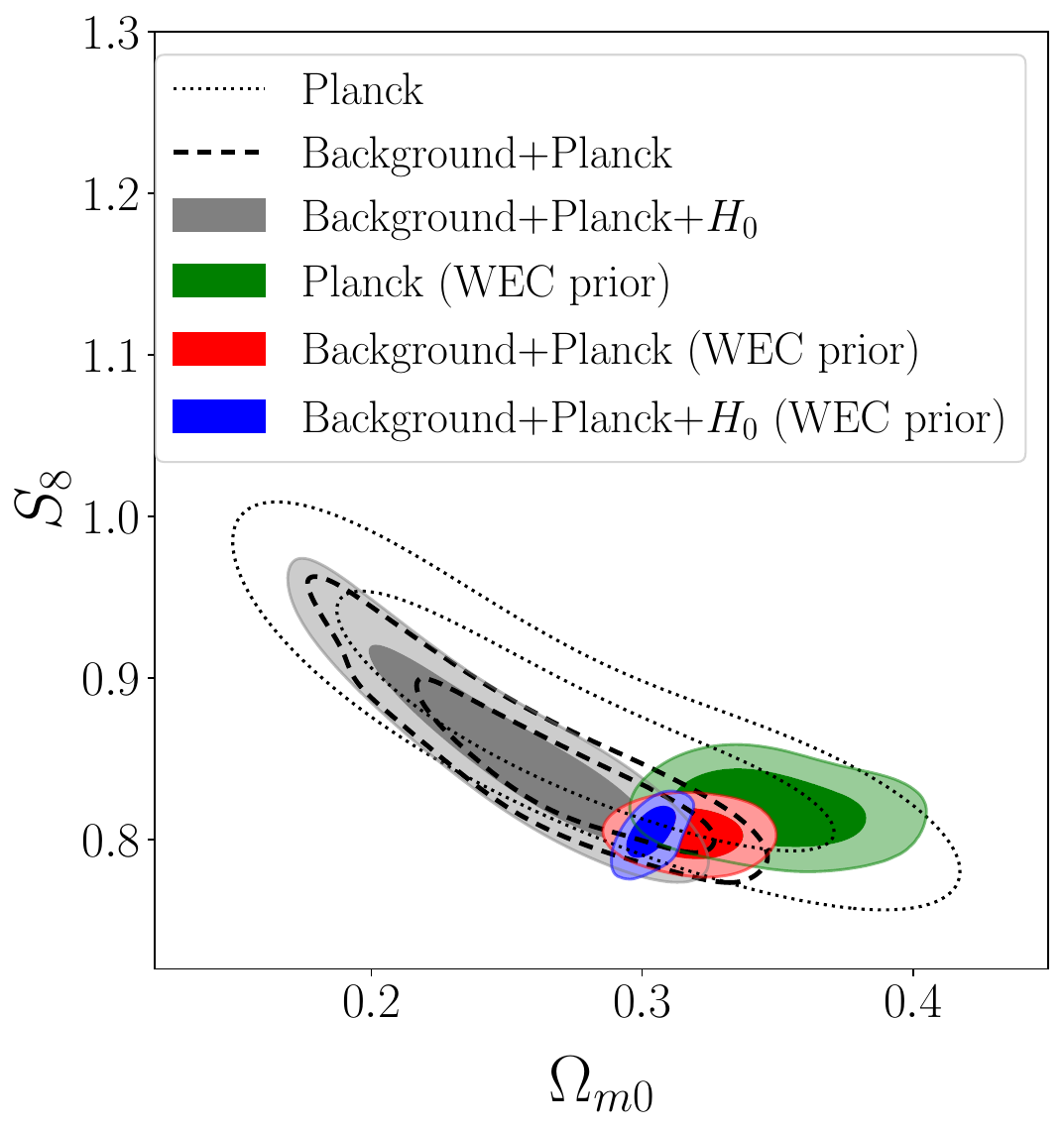}}
\caption{Plot of the matter density parameter $\Omega_{m0}$ versus the parameter $S_8$ without the WEC prior and with the WEC prior. \label{fig:plots_S8}}
\end{figure}

Finally, in Figure \ref{fig:plots_S8}, we present estimates of the quantity $S_8 = \sigma_8 \left({\Omega_{m0}}/{0.3}\right)^{1/2}$, with and without the WEC prior, using the datasets that include Planck. The results of the analyses without the WEC prior have weak constraints for $S_8$ and $\Omega_{m0}$. 
In contrast, the analyses with the WEC prior present good constraints, with mean values around 0.809 and 0.323 for $ S_8$ and $\Omega_{m0}$, respectively. These results are in agreement at $2.3\sigma$ with the Planck 2018 results \cite{Planck-2018vi} and $1.6\sigma$ with the DESI survey \cite{desicollaboration2024desi2024viicosmological} for the $\Lambda$CDM model, and are also consistent with the results of weak gravitational lensing data from KiDS-1000 \cite{Catherine_2021}, as well as clustering and lensing data from the Dark Energy Survey \cite{PhysRevD.98.043528,PhysRevD.105.023514} at $\sim3.5\sigma$ C.L.. The analyses with the WEC prior are also compatible with the results of the analyses for the $\Lambda$CDM model (see \autoref{sec:appA}, Tab. \ref{tab:lcdm_planck}) at $1.3\sigma$ C.L.


\section{Conclusions}\label{sec:Conclusions}
The standard model $\Lambda$CDM presents different problems, such as the cosmological constant problem, the cosmic coincidence, and the Hubble constant tension \cite{Ellis_2011,Sami-2009,Marttens-2014,Weinberg1989,Zlatev1999,VONMARTTENS2019,Peebles2003,Capozziello24,Bamba2012,Pacif2020}, in this sense, several models and approaches have been proposed to better describe the nature of dark energy.
Consequently, several interacting models have been suggested to capture possible interactions between the components of the dark sector of the universe. In this article, we explored an interacting model described by an interacting term proportional to $\rho_x^2/\left(\rho_c+\rho_x\right)$. The direction of the energy transfer depends on the sign of $Q$: when $Q<0$, the process of dark matter creation is enhanced and dark energy decays, whereas when $Q>0$, the opposite occurs. Here, we investigate the theoretical consistency of this class of cosmologies and show that for positive values of $\gamma$ ($\gamma>0$), which physically corresponds to an energy transfer from dark matter to dark energy, this particular model predicts a violation of the WEC, specifically a violation of $\rho_{c}\geq 0$, which will inevitably occur in the future evolution.

The analysis of the results was executed using the CLASS, MontePython, and GetDist codes, employing five different datasets, as described in Section \ref{sec:Methodology}. For all datasets, the results with the WEC prior showed negative values of $\gamma$ at 1$\sigma$ C.L, so excluding the standard $\Lambda$CDM model, which is reproduced by $\gamma=0$.
However, for the 2$\sigma$ C.L. regions, the interaction parameter $\gamma=0$, hence the $\Lambda$CDM model is preferred. 

Our results showed a notable anticorrelation between the interaction parameter $\gamma$ and $\Omega_{m0}$, as well as a correlation between $\gamma$ and the parameters $\sigma_8$, $S_8$, and $H_0$.

In both cases (with and without the WEC), we do not appreciate a significant reduction of the Hubble tension presented in the $\Lambda$CDM model, on the other hand, the inclusion of the WEC prior significantly improves the parameter constraints, showing a preference for drastically lower values of $\sigma_8$, which alleviates the $\sigma_8$ tension between the results from the CMB data (Planck 2018) \cite{Planck-2018vi} and the weak gravitational lensing data (KiDS-1000)  \cite{Catherine_2021}.

The inclusion of the WEC prior significantly improves the parameter constraints, resulting in higher values of $\Omega_{m0}$ and lower values of $\sigma_8$, which combined lead to a lower value of $S_8$, that is included between the values predicted by Planck 2018 \cite{Planck-2018vi}, and the values predicted by the cosmic shear surveys \cite{Catherine_2021,Joudaki_2017,10.1093/mnras/stx2820}.
In any case, our analysis exhibits that the model's predictions are consistent with the current estimates of $\sigma_8$ and $S_8$ within $2\sigma$ C.L. regions with the Planck 2018 results and with the DESI survey \cite{desicollaboration2024desi2024viicosmological}.

\section*{Data Availability Statement}
All data used are already publicly available and properly cited in this paper.

\section*{Acknowledgement}
JSL acknowledges financial support from the Coordenação de Aperfeiçoamento de Pessoal de Nível Superior - Brasil (CAPES) - Finance Code 001.
RvM is suported by Funda\c{c}\~ao de Amparo \`a Pesquisa do Estado da Bahia (FAPESB) grant TO APP0039/2023. LC acknowledges CNPq for the partial support. We acknowledge the use of CLASS, MontePython and \texttt{GetDist} codes.

\bibliographystyle{unsrt}
\bibliography{main}

%

\newpage

\appendix
\section{Observational Results of the $\Lambda$CDM Model}\label{sec:appA}
\begin{figure*}[!htbp]\centering
 \resizebox{0.99\textwidth}{!}{\includegraphics{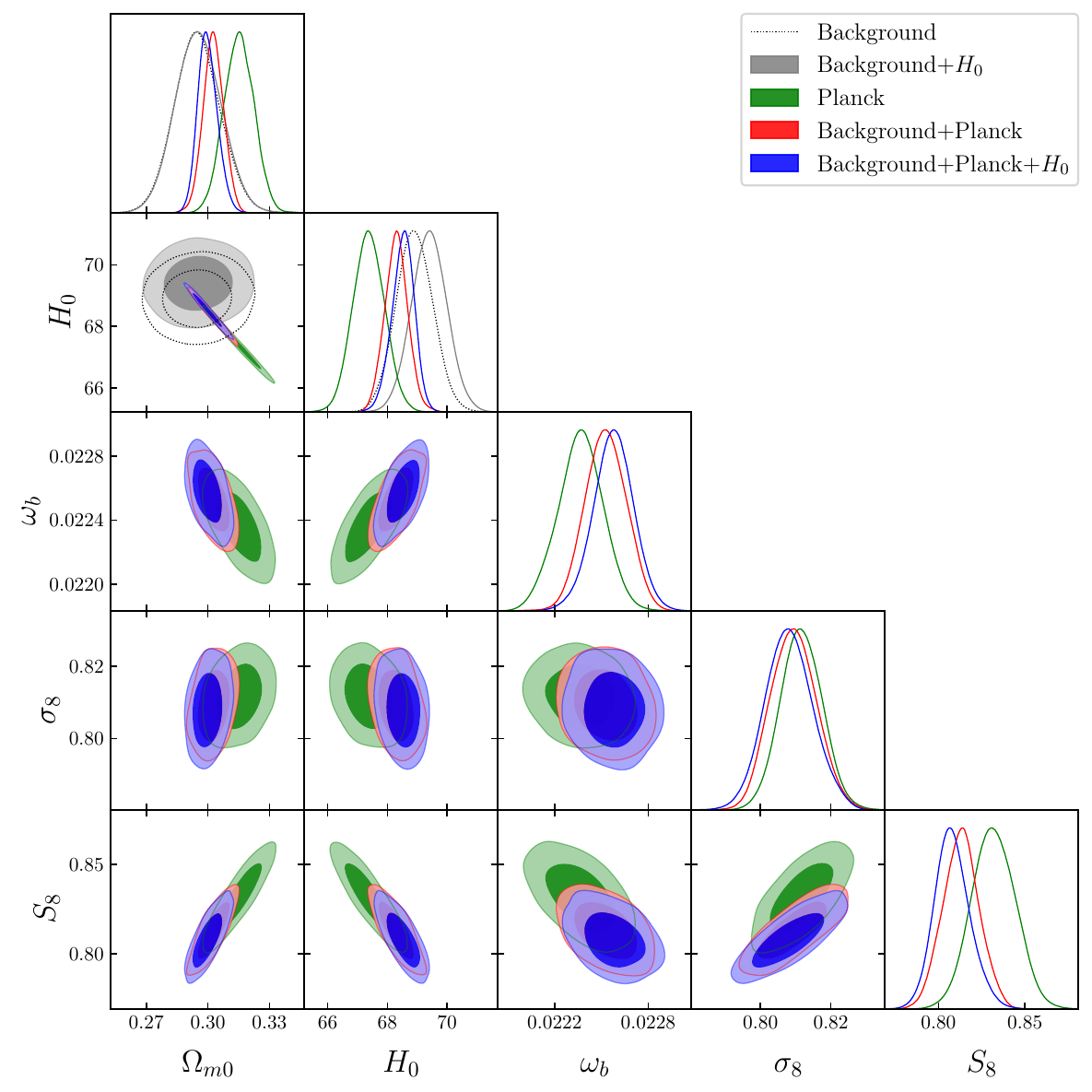}}
\caption{Triangular plot with the free cosmological parameters in the $\Lambda$CDM model, using the datasets: Background, Background+$H_0$, Planck, Background+Planck and Background+Planck+$H_0$. \label{fig:lcdm_retangular}}
\end{figure*}

 Additionally, we will present a triangular plot with the results for the $\Lambda$CDM model, as shown in Fig. \ref{fig:lcdm_retangular}, along with Tables\ref{tab:lcdm_background} and \ref{tab:lcdm_planck}, which display the main cosmological parameters within a \(1\sigma\) C.L. for the \(\Lambda\)CDM model.

\begin{table*}[!htbp]
    \centering
      \caption{Results of the statistical analysis of the $\Lambda$CDM model at $1\sigma$ C.L., using the datasets: Background, and Background+$H_0$.\label{tab:lcdm_background}}
 \begin{tabular} {l c c }
  \\ \toprule \toprule
 Parameter               & Background & Background+$H_0$ \\
\midrule
$\omega_{c}$ & $0.1181^{+0.0056}_{-0.0062}   $ & $0.1203^{+0.0060}_{-0.0060}$\\
$H_0$               & $68.91^{+0.61}_{-0.61}     $ & $69.40^{+0.59}_{-0.59} $ \\\midrule
$\Omega_{m0}$       & $0.295^{+0.011}_{-0.011}$ & $0.295^{+0.011}_{-0.011} $ \\
\bottomrule \bottomrule
\end{tabular}
\end{table*}

\begin{table*}[!htbp]
    \centering
      \caption{Results of the statistical analysis of the $\Lambda$CDM model at $1\sigma$ C.L., using the datasets: Planck, Planck+Background, and Planck+Background+$H_0$.\label{tab:lcdm_planck}}
 \begin{tabular} {l  c c c}
  \\ \toprule \toprule
 Parameter              &  Planck     &  Background+Planck  &  Background+Planck+$H_0$\\
\midrule
{$100\theta{}_{s }$}   & $1.04189^{+0.00029}_{-0.00029}        $ & $1.04209^{+0.00027}_{-0.00027}       $ & $1.04213^{+0.00028 }_{-0.00028}        $\\

{$\ln10^{10}A_{s }$}     & $3.045^{+0.014 }_{-0.014 }           $ & $3.053^{+0.015 }_{-0.015 }            $ & $3.054^{+0.016}_{-0.016}   $\\

{$n_{s }$}    & $0.9650^{+0.0041}_{-0.0041}          $ & $0.9704^{+0.0036}_{-0.0036}$ & $0.9715^{+0.0036}_{-0.0032}          $\\

$\omega_{b}$       & $0.02237^{+0.00014}_{-0.00014}        $ & $0.02253^{+0.00013}_{-0.00013}        $ & $0.02258^{+0.00013}_{-0.00013}         $\\

$\omega_{c}$ & $0.1200^{+0.0012}_{-0.0012}           $ & $0.11803^{+0.00086}_{-0.00086}         $ & $0.11758^{+0.00075}_{-0.00086}         $\\

$\tau_{reio}$& $0.0544^{+0.0074}_{-0.0074}$ & $0.0603^{+0.0075 }_{-0.0075 }          $ & $0.0611^{+0.0067}_{-0.0067}$\\\midrule


$H_0$               & $67.38^{+0.53}_{-0.53}             $ & $68.30^{+0.39 }_{-0.39}  $         & $68.52^{+0.40}_{-0.34}$\\

$\Omega_{m0}$    & $0.3152^{+0.0073}_{-0.0073}          $ & $0.3027^{+0.0050 }_{-0.0050 }         $ &$0.2999^{+0.0043}_{-0.0052}$\\

$\sigma_{8}   $    & $0.8116^{+0.0059}_{-0.0059}          $ & $0.8094^{+0.0064}_{-0.0064}$ & $0.8081^{+0.0067}_{-0.0067}$\\ 
$S_8          $  & $0.832^{+ 0.013  }_{-0.013 } $ & $0.813^{+ 0.010 }_{-0.010} $ & $0.8090^{+0.0091}_{-0.0110}$ \\
\bottomrule \bottomrule
\end{tabular}
\end{table*}

\end{document}